\documentclass[10pt,conference]{IEEEtran}
\pdfoutput=1

\usepackage{layout, calc}
\usepackage{setspace}
\usepackage{times,amssymb,amsmath,epsfig,amsthm}
\usepackage{citesort}

\newcommand{\myincludegraphics}[2]{\includegraphics#1#2}

\newtheorem{theorem}{Theorem}
   \newtheorem{lemma}{Lemma}

   \newtheorem{remark}{Remark}

\newcommand{\bR}{{\mathbf R}}
\newcommand{\bP}{{\mathbf P}}

\title{Diversity Order Gain with Noisy Feedback in Multiple Access Channels }
\author{\authorblockN{Vaneet~Aggarwal}
\authorblockA{
Department of Electrical Engineering\\
Princeton University\\
Princeton, NJ 08544\\
Email: vaggarwa@princeton.edu}
 \and
\authorblockN{Ashutosh~Sabharwal}
\authorblockA{
Department of Electrical \&
Computer Engineering\\
 Rice University\\
Houston, TX 77005\\
Email: ashu@rice.edu} }
\date{}
\begin{document}
\maketitle
\begin{abstract}
%
%
In this paper, we study the effect of feedback channel noise on the diversity-multiplexing tradeoff in multiuser MIMO systems using quantized
feedback, where each user has $m$ transmit antennas and the base-station receiver has $n$ antennas. We derive an achievable tradeoff and use it to show that in SNR-symmetric
channels, a single bit of imperfect feedback is sufficient to
double the maximum diversity order to $2mn$ compared to when there is no feedback (maximum is $mn$ at multiplexing gain of zero). Further, additional feedback bits do not increase this maximum diversity order beyond $2mn$. Finally, the above diversity order gain of $mn$ over non-feedback systems can also be achieved for higher multiplexing gains, albeit requiring more than one bit of feedback.
\end{abstract}

\section{Introduction}

Channel state information to the transmitters has been extensively
studied in MIMO systems
\cite{narula98,Farbod_Dissertation,mukka03,newIT,kim07,gamal06,other,steger,gaj,vaneet,book06,jaf08}
to improve over the diversity-multiplexing tradeoff without
feedback~\cite{zheng03,tse04}). While the earlier work often assumed
noiseless feedback (possibly quantized), recent emphasis has been on
studying the performance with noisy feedback~\cite{steger,gaj,newIT,jaf08}
in single-user MIMO channels. Two distinct models of feedback have
appeared. First is that of two-way training, suitable for symmetric
time-division duplex systems and is the focus of study
in~\cite{steger,gaj}. The other is that of quantized channel state
information~\cite{narula98,Farbod_Dissertation,mukka03,kim07,jaf08} which
is more appropriate for asymmetric frequency-division duplex
systems. In this paper, we study the impact of errors in the
quantized feedback system when used in a multiuser system.


We first model the noise in the feedback which depends on the
signal-to-noise ratio of the channel from the transmitter to the
receiver. Thus, we visualize the feedback channel also operating
over a noisy communication link.
Next, we bound the probability of outage to derive the
diversity-multiplexing performance of MAC (Multiple Access Channel) for the noisy feedback
model for quantized channel state information. The general result leads to the following conclusions
about multiple access and as special case, single-user MIMO channels.

If the forward and feedback channel are SNR-symmetric (true if the
nodes have identical power constraints operating over statistically
identical channels), then feedback errors limit the maximum
diversity order to $2mn$, achieved at zero multiplexing point. The
diversity order of $2mn$ is double that of what can be achieved with
no feedback~\cite{zheng03,tse04} and is identical to that achieved
by two-way training method studied in~\cite{steger,gaj}. Thus the
two dominant models achieve the same maximum diversity order when
the transmitter is mismatched with the receiver, thus providing a
satisfying conclusion. At the same time, it is interesting to note
that a single noisy bit of information is same as training the full
channel from the point of view of diversity-multiplexing tradeoff.

However, the picture changes for higher multiplexing gains. While
there is no way of controlling diversity order gain with two-way
training of~\cite{steger,gaj}, more feedback bits lead to different
level of diversity order gains. We also show that as the number of
feedback bits grow, the diversity order of $\left( mn +
\text{diversity order achievable without feedback}\right)$ can be
achieved. For example, in  single user MIMO channel, a diversity
order of $mn + (m-r)(n-r)$ for integer multiplexing gains $0\le r<
\min(m,n)$ can be achieved with finite number of error-prone
feedback bits, a number which we quantify.

We highlight the fact that all our results are derived for a
multiuser system with $L$ users, each with $m$ transmit antennas and
a receiver with $n$ receive antennas. This in contrast to most of
the earlier work which has considered noisy channel state feedback
in the context of single user
systems~\cite{steger,newIT,other,book06,jaf08}.

%

%

%


The rest of the paper is outlined as follows. In Section II we give
background on the channel model, introduce feedback model and
diversity-multiplexing tradeoff. In Section III, we find the
diversity-multiplexing tradeoff for Multiple Access Channel. In
Section IV, we discuss these results. Section V concludes the paper.

\section{System Model}
\subsection{Channel Model}
Consider a multiple access channel with $L$ users where the
transmitters have an array of $m$ transmit antenna and the receiver
has an array of $n$ receive antenna. The channel is constant during
a fading block of $T$ channel uses, but changes independently from
one block to the next. During a fading block $l$, the channel is
represented by $n \times m$ random matrices $H_{s,l}$ ($1\le s \le
L$), and the received signal can be written in the matrix form as $
Y_l = \mathop \sum \limits_{1 \le i \le L} H_{i,l}X_{i,l}+ W_l.$
Here, $W_l$ of size $n \times T$ represents additive white Gaussian
noise at the receiver with all entries i.i.d. $CN(0,1)$. We consider
a richly scattered Rayleigh fading environment, i.e. elements of
$H_{s,l}$ are assumed to be i.i.d $CN(0,1)$. The transmitters are
subject to an average power constraint such that the long-term power
is upper bounded, i.e, ${\mathbb E}\left[ X_{s,l}^2 \right] \leq
\mathsf{SNR}_i$ for $1\le s\le L$.

\subsection{Feedback Model \label{sec:feedback model}}

We will assume that the receiver has perfect knowledge of the
channel coefficients $H_{s,l}$ ($1\le s \le L$). We denote ${\mathbf
H}_l = (H_{1,l},H_{2,l}, ... H_{L,l})$. The receiver then uses the
knowledge of channel coefficients to compute a feedback signal
${\rm I}({\mathbf H}_l)$ which is sent to the transmitters. Furthermore,
we will assume that this feedback signal takes on only finite number
of values from the set $\{1,2,\ldots,K\}$, where $K>1$. Note that
when $K=1$, there is no feedback, and hence the case reduces to that
in \cite{tse04}. Finally, the mapping ${\rm I}({\mathbf H}_l): {\mathbf
H}_l \mapsto \{1,2,\ldots,K\}$ is a deterministic function which can
potentially depend on the $\mathsf{SNR}$ and the rate of
transmission. Due to the error in the feedback, the users do not
receive the same signal as is sent by the receiver. The feedback
channel is modeled as follows. Let ${\rm I}({\mathbf H}_l)=i$ be
transmitted from the receiver. User $s$ receives an index
$\overline{\rm I}_s$ which takes on only finite number of values from
the set $\{1,2,\ldots,K\}$ and is given by
\vspace{-.06in}
\[
\overline{\rm I}_s = \left\{ \begin{array}{l}
  i  \text{ with probability } 1-\epsilon\\
  i' \ne i \text{ with probability } \frac{\epsilon}{K-1}\\
 \end{array} \right. \text{ for } 1\le s \le L,
\]
where $\epsilon$ depends on $\mathsf{SNR}$. 
\subsection{Diversity-Multiplexing Tradeoff Definitions}\label{dmmimo}
A codeword $X_{s,l}$ is assumed to span a single fading block. Since
we do not consider coding over multiple fading blocks, the block
index $l$ will be omitted whenever this does not cause any
confusion. Conditioned on indices $\overline{\rm I}_s=i_s$, the
transmitter $s$ chooses a codeword $X_s$ from the codebook
$C_{s,i_s} = \{X_{s,i_s}(1), X_{s,i_s}(2), ... , X_{s,i_s}(E_{s})\}$
of rate $R_s$ for $1\le s \le L$. All the $X_{s,i}(k)$'s are
matrices of size $m \times T$.

In this paper, we will only consider single rate transmission where
the rate of the codebooks does not depend on the feedback index and is known to the receiver. Therefore, regardless of which feedback index the transmitters receive, the receiver
attempts to decode the received codeword from the same codebook. Outage occurs when the transmission
power is less than the power needed for successful (outage-free) transmission.

The average power constraint at each transmitter can be given along the lines of \cite{kim07} as follows. First
define average power per codeword
\vspace{-.1in}
\[
P_s^i \triangleq \frac{1}{T E_{s}}\sum\limits_{k=1}^{E_{s}}
||X_{s,i}(k)||_F^2 \ \ , 1\le s \le S
\]\vspace{-.01in}
which leads to average power constraints
\vspace{-.09in}
\begin{eqnarray}
E_{{\mathbf H}}[P_s^{\overline{\rm I}_s({\mathbf H})}] \triangleq
\sum\limits_{i = 1}^K {P_s^i \Pi (\overline{\rm I}_s({\mathbf H} ) = i)}
\leq \mathsf{SNR}_s,  1\le s\le L\label{eq:power_constraint}
\end{eqnarray}
where $\Pi(\alpha)$ denotes probability of event $\alpha$.

Since our focus is asymptotic
performance behavior in the form of diversity-multiplexing tradeoff,
we will assume that $\mathsf{SNR}_s  \buildrel.\over= \mathsf{SNR}$ for all $1\le s \le
L$.\footnote{We adopt the notation
of \cite{zheng03} to denote $ \buildrel.\over=$ to represent
exponential equality. We similarly use $ \buildrel.\over<$, $
\buildrel.\over>$, $ \buildrel.\over\le$, $ \buildrel.\over\ge$ to
denote exponential inequalities.} Note that all the index mappings, codebooks, rates, powers are
dependent on $\mathsf{SNR}$s. The dependence of rates on the
$\mathsf{SNR}$s is explicitly given by $R_s = r_s \log \mathsf{SNR}_s$.
We refer to ${\mathbf r}\triangleq (r_s)_{1\le s \le L}$ as the
multiplexing gains.

In point-to-point channels, outage is defined as the event that the
mutual information of the channel, $I(X;Y | H)$ is less than the
desired rate $R$, where $I(X;Y|H) = \log\det\left(I+\frac{P}{m}HQH^\dagger\right)$ is the mutual information of a
point-to-point link with $m$ transmit and $n$ receive
antennas, transmit power $P$ and input distribution Gaussian with covariance matrix $Q$~\cite{zheng03}. Since $I(X;Y|H)$ depends on transmit power, we write this dependence explicitly as $I(X;Y | H,P)$. In a multiple access channel,
corresponding outage event is defined as the event that the channel
cannot support target data rate for all the users~\cite{tse04}.

Hence, for a multiple access channel with $L$ users, each equipped
with $m$ transmit antennas, and a receiver with $n$ receive
antennas, the outage event is $O({\mathbf R},{\mathbf P}) \triangleq \mathop \cup \limits_S
O_S({\mathbf R},{\mathbf P})$ where ${\mathbf P}=(P_1,P_2,.., P_L)$ and ${\mathbf R}=(R_1,R_2,...,R_L)$. The union is
taken over all subsets $S \subseteq \{1,2, ..., L\}$, and
$O_S({\mathbf R}, {\mathbf P}) \triangleq \{{\mathbf H} \in \Bbb C^{n\times Lm} : I(X_S;Y |
X_{S^c},{\mathbf H},{\mathbf P}) < \mathop \sum \limits_S R_i \}$ where $X_S$ contains
the input signals from the users in $S$ with powers ${\mathbf
P}$. As before, $I(X_S;Y | X_{S^c}, {\mathbf H},\mathbf{P})$ represents $I(X_S;Y | X_{S^c}, {\mathbf H})$ when the transmit powers are ${\mathbf P}$. Let $\Pi(O)$ denote the probability of outage. The
system is said to have diversity order of $d$ if
$\Pi(O) \doteq \mathsf{SNR}^{-d}$. The diversity multiplexing for
the multiple users can be described as: given the multiplexing gains
${\mathbf r}$ for all the users, the diversity order that  can be
achieved describes the diversity-multiplexing tradeoff region.

The probability of outage with rate ${\mathbf R}=(R_1,R_2, ...,
R_L)$ and transmit power ${\mathbf P}=(P_1,P_2, ..., P_L)$ is
denoted by $\Pi({\mathbf R},{\mathbf P})\triangleq \Pi(O({\mathbf R},{\mathbf P}))=\Pi(\mathop \cup \limits_S
O_S({\mathbf R}, {\mathbf P}))$. Also let
$U({\mathbf R},{\mathbf P})$ be defined as the indicator function of
$\mathop \cup \limits_S
O_S({\mathbf R}, {\mathbf P})$.
Then, $\Pi({\mathbf R},\mathbf {P})$ is the probability of event
$\{U({\mathbf R},{\mathbf P})=1\}$ over the randomness of channel
matrices. Let $P_s\buildrel.\over= \mathsf{SNR}^{p_s}$ for all $1\le s \le
L$. Further let ${\mathbf R} = (R_1,R_2, ..., R_L) \doteq (r_1\log
\mathsf{SNR}, r_2 \log \mathsf{SNR}, ..., r_L \log \mathsf{SNR})$. Let $D({\mathbf
r},{\mathbf p})$ be defined as $\Pi({\mathbf
R}, {\mathbf P}) \buildrel.\over= \mathsf{SNR}^{-D({\mathbf
r},{\mathbf p})}$ where ${\mathbf r}=(r_1,r_2, ..., r_L)$ and ${\mathbf
p}=(p_1,p_2, ..., p_L)$.
\vspace{-.08in}

\begin{lemma}\label{outlemma}
Let $p_s=p$ for all $1\le s \le
L$.  Also, let $\mathop \sum \limits_{i \in S}r_i\le \min(|S|m,n)$ for
all non-empty subsets $S$ of $\{1, 2, ..., L\}$. Then, \vspace{-.1in}
\begin{eqnarray}
D({\mathbf r}, {\mathbf p}) = \mathop {\min}
\limits_{S}G_{|S|m,n}\left(\mathop \sum \limits_{i \in S}r_i,\mathop
p\right)
\end{eqnarray}\vspace{-.12in}
\noindent where $G_{m,n}(r,p) \triangleq $
\[
 \inf \limits_{\alpha_1^{\min(m,n)}\in A}
\mathop \sum
\limits_{i=1}^{\min(m,n)}(2i-1+\max(m,n)-\min(m,n))\alpha_i
\]
with $A \triangleq $
\[
\{\alpha_1^{\min(m,n)} | \alpha_1 \ge \ldots \alpha_{\min(m,n)}
\ge 0, \mathop \sum \limits_{i=0}^{\min(m,n)}(p-\alpha_i)^+<r\}.
\]
\end{lemma}\vspace{-.2in}
\noindent {\bf Proof}: Note that  $\Pi ({\mathbf R},{\mathbf P}) =  {\Pi}_{\mathbf H}
\left(\bigcup\limits_S {O_S ({\mathbf R},{\mathbf P})} \right)$. Hence,
${\Pi}_{{\mathbf H}} (O_S ({\mathbf R},{\mathbf P}) \leqslant \Pi ({\mathbf
R},{\mathbf P}) \leqslant \sum\limits_S {{\Pi} _{\mathbf H} (O_S ({\mathbf R},{\mathbf
P}))}$. As, $\Pi_H (O_S ({\mathbf R},{\mathbf
P}))$ is probability of outage for single
user with $|S|m$ transmit antennas, $n$ receive antennas, rate
$\doteq \sum\limits_{i \in S} {r_i}\log(\mathsf{SNR})$, power
$\doteq \mathsf{SNR}^{p}$, by
\cite{kim07}, ${\Pi} _{\mathbf H} (O_S ({\mathbf R}, {\mathbf P}))
 \doteq \mathsf{SNR}^{ - G_{|S|m,n}
(\sum\limits_{i \in S} { r_i } ,p
)}. $
Hence,  $\Pi({\mathbf R}, {\mathbf P}) \buildrel.\over=
\mathsf{SNR}^{-D({\mathbf r},{\mathbf p})}. \quad \blacksquare$
\vspace{-.1in}
\begin{remark} {\rm
$G_{m,n}(r,p)$ is a piecewise linear curve connecting the points $(r,G_{m,n}(r,p))$=
$(kp,p(m-k)(n-k))$, $k = 0, 1, \ldots, \min(m,n)$ for fixed $m$, $n$ and $p>0$. This follows
directly from Lemma 2 of \cite{kim07}.}\end{remark}
\vspace{-.15in}
\subsection{Feedback-based Power Control}
\vspace{-.05in}
In this section, we describe the power control policy for the
optimum receiver for which successful decoding occurs if the transmission power is greater than or equal to the power needed for outage-free transmission.
Recall that the sent feedback signal ${\rm I}$ and the
received feedback signal $\overline{\rm I}_s$ takes values over a finite
set as described in Section~\ref{sec:feedback model}. For each
received index $\overline{\rm I}_s=\overline{i}_s$ at User $s$, the
transmitted power is denoted by $P_s^{\overline{i}_s}$. We assume
that $P_s^1 \leq P_s^2 \leq \cdots \leq P_s^K$. We denote the power
tuple as ${\mathbf P}_{i} = (P^{i}_1,P^{i}_2, ..., P_L^i)$.
Following \cite{kim07,vaneet}, ${\rm I}=i$ is calculated for as\vspace{-.05in}
\[
i = \left\{ \begin{array}{l}
  1  \text{ if }U({\mathbf R},{\mathbf P}_{K})=1\\
  \min_{ k \in \{1, \ldots , K\}} \{U({\mathbf R},{\mathbf P}_{k})=0\} \text{ otherwise}\\
 \end{array} \right. .
\]
According to the scheme, we transmit at minimum power level needed for outage-free transmission in case outage can be avoided, and send at minimum power level in case it cannot be avoided.
Using the scheme, we can compute the probability of occurrence of
event (${\rm I}=i$) as
\vspace{-.1in}
\begin{equation}\label{proi}
\Pi({\rm I}=i) = \begin{cases}
1+ \Pi(\bR,\bP_{K}) - \Pi(\bR,\bP_{1}), & i = 1 \\
\Pi(\bR,\bP_{i-1}) - \Pi(\bR,\bP_{i}), & 2 \leq i \leq K
\end{cases}.\end{equation}
The power levels are chosen to minimize the outage probability
$\Pi(O) $ subject to the power
constraint~(\ref{eq:power_constraint}).
 \vspace{-.05in}
\section{Diversity-Multiplexing Tradeoff}
In this section, we will give an achievable diversity multiplexing tradeoff with errors in feedback with certain cases when this is the best achievable. We assume that the feedback errors decays with $\mathsf{SNR}$ as $\epsilon \doteq \mathsf{SNR}^{-y}$.

When $K=1$, there is no feedback and hence no imperfection. The
diversity for any multiplexing vector ${\mathbf r} = (r_1,r_2, ..,
r_L)$ is given by $D({\mathbf r},{\mathbf 1})$ for
$\mathop\sum\limits_{i \in S}r_i<\min(|S|m,n)$ for all non-empty
subsets $S$ of $\{1, 2, ..., L\}$ where ${\mathbf 1}$ is a vector of
length $L$ containing all ones~\cite{tse04}. So, we only consider
the case $K>1$ in this section.

 Let $C_{m,n,K}({\mathbf
r})$ be given by a recursive equation
\begin{eqnarray}\label{ceq}
C_{m,n,j} ({\mathbf r} ) = \left\{ \begin{array}{l}
 0\text{ when $j = 0$} \\
 D({\mathbf r} ,{\mathbf 1}(1+C_{m,n,j-1}({\mathbf r})) ) \text{ when $j \ge 1$} \\
 \end{array} \right. .\nonumber
\end{eqnarray}
\begin{theorem} \label{thrm:macl1kg1}Suppose that $K>1$ and $\epsilon\doteq \mathsf{SNR}^{-y}$ for some
$y>0$. Further suppose that $\mathop\sum\limits_{i \in
S}r_i<\min(|S|m,n)$ for all non-empty subsets $S$ of $\{1, 2, ...,
L\}$. Then, the lower bound for diversity-multiplexing tradeoff  is given by
$d_{opt}^{K}=\min(\overline{C}_{m,n,K}({\mathbf r}),
y+C_{m,n,1}({\mathbf r}))$ where $\overline{C}_{m,n,K}({\mathbf r})$
is given by a recursive equation
\begin{eqnarray}\label{ceq}
\overline{C}_{m,n,j} ({\mathbf r} ) = \left\{ \begin{array}{l}
 0\text{ when $j = 0$} \\
 D({\mathbf r} ,{\mathbf 1}(1+\min(y,\overline{C}_{m,n,j-1}({\mathbf r}))) ) \text{ when $j \ge 1$} \\
 \end{array} \right. \nonumber
\end{eqnarray}
\end{theorem}
\noindent {\bf Proof}: The probability of outage for this scheme can be bounded as\vspace{-.1in}
\begin{eqnarray}\label{poutl1kg1}\Pi(O) &\le&\Pi({\mathbf R},{\mathbf
P}_{K})+\sum_{s=1}^L\sum_{i=2}^K
\Pi(\overline{I}_s<i|I=i)\Pi(I=i)\nonumber\\
&=&(1-L\epsilon)\Pi({\mathbf R},{\mathbf
P}_{K})+\frac{L\epsilon}{K-1}\sum_{i=1}^{K-1}{\Pi({\mathbf
R},{\mathbf P}_{i})}
\end{eqnarray}
We next calculate the probability that $\overline{\rm I}_s=i$ as
\begin{equation}\label{i12overlinel1kg1}\Pi(\overline{\rm I}_s=i)=\frac{\epsilon}{K-1}+\left(1-\frac{\epsilon
K}{K-1}\right)\Pi({\rm I}=i).\end{equation}

The power levels are selected to minimize outage probability subject
to power constraints \eqref{eq:power_constraint}.

Consider the power levels as:
\[
\widetilde P_s^{i}  = \left\{ \begin{array}{l}
 \frac{\mathsf{SNR}_s}{K}\text{ when $i = 1$} \\
 \frac{\mathsf{SNR}_s}{{K (\frac{\epsilon}{K-1}+(1-\frac{\epsilon K}{K-1})\Pi({\mathbf R} ,\widetilde {\mathbf P}_{i-1} ))}}\text{ when $i > 1$} \\
 \end{array} \right. \forall 1\le s\le L.
\]

These power levels satisfy the $\mathsf{SNR}$ constraints, and hence
the optimal outage probability is $\le$ the outage probability with
these power levels. Let $\widetilde{\Pi}(O)$ be the outage
probability using power levels $\widetilde P_s^{i}$. Then, $\Pi(O)
\buildrel.\over \le \widetilde{\Pi}(O)$.

From these power levels, we find that\vspace{-.05in}
\begin{eqnarray}
\widetilde P_s^{i}  \doteq
\mathsf{SNR}^{1+\min(y,\overline{C}_{m,n,i-1}({\mathbf r}))}
\end{eqnarray}
Hence,\vspace{-.2in}
\begin{eqnarray}
\widetilde{\Pi}(O) &\buildrel.\over \le&(1-L\epsilon)\Pi({\mathbf
R},\widetilde{\mathbf
P}_{K})+\frac{L\epsilon}{K-1}\sum_{i=1}^{K-1}{\Pi({\mathbf
R},\widetilde{\mathbf P}_{i})} \nonumber\\&\doteq&
\mathsf{SNR}^{-\overline{C}_{m,n,K}({\mathbf r})}+
\mathsf{SNR}^{-y-\overline{C}_{m,n,1}({\mathbf r})}\nonumber\\
&\doteq& \mathsf{SNR}^{-\min(\overline{C}_{m,n,K}({\mathbf r}),
y+\overline{C}_{m,n,1}({\mathbf r}))}\nonumber
\end{eqnarray}
Hence, we find that the outage probability is $\buildrel.\over \le
\mathsf{SNR}^{-\min(\overline{C}_{m,n,K}({\mathbf r}),
y+\overline{C}_{m,n,1}({\mathbf r}))}$. Noting that
$\overline{C}_{m,n,1}({\mathbf r}) =C_{m,n,1}({\mathbf r})$ proves
the theorem.\quad $\blacksquare$

Theorem \ref{thrm:macl1kg1} gives a lower bound to the
diversity-multiplexing tradeoff performance. We will now consider
some special cases when this bound is tight.

\begin{lemma}\label{l1kg1le} Suppose that $K>1$, $\epsilon=0$ and $\mathop\sum\limits_{i \in S}r_i<\min(|S|m,n)$ for all non-empty
subsets $S$ of $\{1, 2, ..., L\}$. Then, the diversity-multiplexing
tradeoff  with $K$ indices of global
feedback is given by $C_{m,n,K}({\mathbf r})$ for given multiplexing
gain ${\mathbf r} = (r_1,r_2, ..., r_L)$.
\end{lemma}
\noindent {\bf Proof}: The lower bound of the diversity multiplexing tradeoff follows from
Theorem~\ref{thrm:macl1kg1} by taking limit as $y\to \infty$.  We will now
prove the upper bound for the diversity multiplexing tradeoff by
finding the lower bound for outage probability.

Note that $\Pi(O)= \Pi({\mathbf R}, {\mathbf P}_K)$ when there is no
error in the feedback. To calculate the lower bound for outage
probability, we first weaken the above optimization problem as $\min
\Pi(O)$ subject to the following power constraint
\begin{eqnarray}
 {\Pi(I=i)P_s^i} \le \mathsf{SNR}_s\text{ } \forall 1 \le s \le L\label{eqweeka}
\end{eqnarray}
The solution of (\ref{eqweeka}) is denoted by $\overline{ P}_s^{i}$.
As the constraint set is bigger compared to the original problem, it
follows that $\Pi(O) \buildrel.\over \ge \overline{\Pi}(O)$ where
$\overline{\Pi}(O)$ is the outage probability taking powers
$\overline{ P}_s^{i}$. Note from \eqref{eqweeka} that
$\overline{P}_s^1 \le K \mathsf{SNR}_s$ which gives
$\overline{P}_s^1 \buildrel .\over \le \mathsf{SNR}$. Using
\eqref{eqweeka} and \eqref{proi} recursively, we find that \vspace{-.2in}

\begin{eqnarray}
\overline P_s^{i}  \buildrel .\over \le
\mathsf{SNR}^{1+{C}_{m,n,i-1}({\mathbf r})}
\end{eqnarray}
Hence, the outage probability
\begin{eqnarray}\label{lowerbd}
\Pi(O)\nonumber &\buildrel.\over \ge&\Pi({\mathbf
R},\overline{\mathbf P}_K)\nonumber\\&\buildrel.\over \ge&
\mathsf{SNR}^{D({\mathbf
r},\mathbf{1}(1+{C}_{m,n,K-1}({\mathbf r})))}\nonumber \\
&\buildrel.\over =&\mathsf{SNR}^{-{C}_{m,n,K}({\mathbf r})} \quad \blacksquare
\end{eqnarray}
\vspace{-.3in}
\begin{remark}{\rm Lemma \ref{l1kg1le} has earlier been proved in \cite{kim07} for
$L=1$ and \cite{vaneet} for $L=2$.}
\end{remark}

\begin{lemma}\label{l1lemma}
Suppose that $K>1$, $L=1$ and $\epsilon\doteq \mathsf{SNR}^{-y}$ for
some $y>0$. Further suppose that $r_1<\min(m,n)$. Then, the
diversity-multiplexing tradeoff for the optimal receiver with $K$
indices of feedback is given by
$d_{opt}^{K}=\min(\overline{C}_{m,n,K}({\mathbf r}),
y+C_{m,n,1}({\mathbf r}))$.
\end{lemma}
\noindent {\bf Proof}: The lower bound follows by Theorem \ref{thrm:macl1kg1}. We will now
prove the upper bound for the diversity multiplexing tradeoff by
finding the lower bound for outage probability.

For this, we first weaken the above optimization problem as $\min
\Pi(O)$ subject to the following power constraint
\begin{eqnarray}
 \frac{\epsilon}{K-1} P_1^i+\left(1-\frac{\epsilon
K}{K-1}\right){\Pi(I=i)P_1^i} \le \mathsf{SNR}_1 \label{eqweek}
\end{eqnarray}
The solution of (\ref{eqweek}) is denoted by $\overline{ P}_1^{i}$.
As the constraint set is bigger compared to the original problem, it
follows that $\Pi(O) \buildrel.\over \ge \overline{\Pi}(O)$ where
$\overline{\Pi}(O)$ is the outage probability taking powers
$\overline{ P}_1^{i}$.

Note that $\overline{P}_1^1 \le K \mathsf{SNR}$. From
\eqref{eqweek}, it follows that $\frac{\mathsf{SNR}}{\overline
P_1^j} \ge \frac{\epsilon}{K-1}+\left(1-\frac{\epsilon
K}{K-1}\right)\Pi(I=j)$. Hence, $\overline P_1^j \buildrel.\over\le
\mathsf{SNR}^{1+y}.$ Assuming that $\overline P_1^j \buildrel.\over
\le \mathsf{SNR}^{1+y}$, we find using \eqref{eqweek} that
$\overline P_1^j \buildrel.\over\le
\mathsf{SNR}^{1+G_{m,n}(r,\overline p_{j-1})}$, for $\overline
P_1^{j-1} \buildrel.\over\le \mathsf{SNR}^{\overline p_1^{j-1}}$.
Using this recursively, we find that $\overline P_1^j
\buildrel.\over\le \mathsf{SNR}^{1+\min(y,\overline
C_{m,n,j-1}(r))}$ and hence $\Pi({\mathbf R},\overline P_1^j)
\buildrel.\over\ge \overline{C}_{m,n,j}({\mathbf r}) $.

Also, since $L=1$,\vspace{-.1in}
\begin{eqnarray}
\label{poutl1}\Pi(O) &=&\Pi({\mathbf R},{\mathbf
P}_{K})+\sum_{i=2}^K \Pi(\overline{I}_1<i|I=i)\Pi(I=i)\nonumber\\
&=&(1-\epsilon)\Pi({\mathbf R},{\mathbf
P}_{K})+\frac{\epsilon}{K-1}\sum_{i=1}^{K-1}{\Pi({\mathbf
R},{\mathbf P}_{i})}
\end{eqnarray}

Hence, the outage probability $\Pi(O)$ \vspace{-.1in}
\begin{eqnarray}\label{lowerbd}
\nonumber &\buildrel.\over \ge&
(1-\epsilon)\mathsf{SNR}^{-\overline{C}_{m,n,K}({\mathbf
r})}+\frac{\epsilon}{K-1}\sum_{i=1}^{K-1}{\mathsf{SNR}^{-\overline
C_{m,n,i}({\mathbf r})}}\nonumber \\
&\buildrel.\over =&\mathsf{SNR}^{-\min(\overline{C}_{m,n,K}({\mathbf
r}), y+\overline{C}_{m,n,1}({\mathbf r}))}\quad \blacksquare
\end{eqnarray}
\vspace{-.35in}
\begin{remark}{\rm It was recently observed in \cite{jaf08} that for $K>1$ and $L=1$, we do not gain in diversity order with feedback if the feedback errors do not decay with $\mathsf{SNR}$. This also follows as a special case of Lemma \ref{l1lemma} with $y\to0$ in which case, $d_{opt}^{K}=C_{m,n,1}({\mathbf r})$ is the same as the diversity order without feedback.}
\end{remark}\vspace{-.25in}
\section{Doubling of Diversity Order}

Theorem \ref{thrm:macl1kg1} gives achievable diversity-multiplexing
tradeoff for MAC with imperfect feedback. Note that the theorem
considered $\epsilon \doteq \mathsf{SNR}^{-y}$ for any $y>0$. We saw
in Lemma \ref{l1kg1le} the performance of MAC with perfect feedback.
When the feedback error does not decay with $\mathsf{SNR}$, we get
$d_{opt}^K= C_{m,n,1}({\mathbf r})= D({\mathbf r}, {\mathbf 1})$.
This is same as the diversity-multiplexing tradeoff without feedback
\cite{tse04} and hence if the feedback error does not decay with
$\mathsf{SNR}$, the feedback do not help in getting any increase in diversity order.

When the forward and the backward channel are
$\mathsf{SNR}$-symmetric, the feedback error from the
transmitter to the receiver scales as $\epsilon \doteq
\mathsf{SNR}^{-mn}$.\footnote{Note that the receiver which is sending back the feedback is assumed to operate without any channel state information, especially when operating in an FDD system.} Thus, $y=mn$. Next, we analyze the performance
loss with imperfection in feedback.

\begin{lemma}[Doubling of Diversity Order]
\label{limle} Let $y=mn$. When ${\mathbf r}\to \mathbf{0}$, the
diversity order is given by
\begin{eqnarray}
d_{opt}^K  = \left\{
\begin{array}{l}
 mn\text{ when $K = 1$} \\
 2mn \text{ when $K > 1$} \\
 \end{array} \right. .\nonumber
\end{eqnarray}
Furthermore, as ${\mathbf r}\to \mathbf{0}$,  $C_{m,n,K}$ and $\overline{C}_{m,n,K}$ behave as
\begin{eqnarray}
C_{m,n,K}({\mathbf 0})=mn\left(\frac{(mn)^K-1}{mn-1}\right) \text{
for } K\ge 1. \nonumber
\end{eqnarray}
\begin{eqnarray}
\overline{C}_{m,n,K} ({\mathbf 0} ) = \left\{ \begin{array}{l}
 mn\text{ when $K = 1$} \\
 mn(1+mn) \text{ when $K > 1$} \\
 \end{array} \right. .\nonumber
\end{eqnarray}
\end{lemma}

When the feedback is perfect, diversity order increases exponentially with the number of feedback indices \cite{vaneet} while
Lemma \ref{limle} shows that if the feedback
is imperfect, the diversity order do not increase with feedback (for
$K>1$). Lemma \ref{limle} also shows that diversity order of $2mn$
is achievable as multiplexing gains go to zero for any number of
indices of feedback ($K>1$), and hence also for single-bit of
feedback. This further means that achievable diversity order doubles with just a
single-bit of feedback compared to the case of no feedback for zero
multiplexing gains.

In \cite{steger,gaj}, it was shown for MIMO channels that diversity
of $2mn$ can be achieved by training. In this paper, by Lemma
\ref{limle}, we have shown that diversity order of $2mn$ can be
achieved with just a single bit of feedback. Hence, the training can
be replaced with just a single bit of feedback if the objective is
just to achieve diversity of $2mn$ for zero multiplexing gains.

Till now, we focussed on zero multiplexing gains. Now, we consider
general multiplexing gains satisfying $\mathop\sum\limits_{i \in
S}r_i<\min(|S|m,n)$.
\begin{lemma}\label{hel} Let ${\mathbf r}= (r_1, ..., r_L)$ with $\mathop\sum\limits_{i \in
S}r_i<\min(|S|m,n)$. Then, the following holds:
\begin{eqnarray}
D({\mathbf r},{\mathbf 1}(1+mn)) \ge mn+D({\mathbf r},{\mathbf 1})
\end{eqnarray}
\end{lemma}
\noindent {\bf Proof}: We will be done if we prove that $G_{|S|m,n}(\mathop \sum
\limits_{i\in S}r_i,1+mn)\ge mn+G_{|S|m,n}(\mathop \sum
\limits_{i\in S}r_i,1)$. Hence, it is enough need to prove that
$G_{m,n}(t,1+mn)\ge mn+G_{m,n}(t,1)$ for $t< \min(m,n)$. Note that $G_{m,n}(\lceil t\rceil,1+mn)= mn(1+mn)-\lceil
t\rceil(m+n-1)\ge mn+(m-\lceil t\rceil)(n-\lceil
t\rceil)=mn+G_{m,n}(\lceil t\rceil,1)$. Similarly, $G_{m,n}(\lfloor
t\rfloor,1+mn)\ge mn+G_{m,n}(\lfloor t\rfloor,1)$. Since, both
$G_{m,n}(x,1+mn)$ and $mn+G_{m,n}(x,1)$ are linear in $x$ for
$\lfloor t\rfloor \le x \le\lceil t\rceil$, the result follows.\quad $\blacksquare$

Note that $C_{m,n,j+1}({\mathbf r})>C_{m,n,j}({\mathbf r})$ for
$\mathop\sum\limits_{i \in S}r_i<\min(|S|m,n)$. Also,
$C_{m,n,1}({\mathbf r})\le mn$. Hence, there exist a $k\ge 1$ such
that:
\begin{eqnarray}
{C}_{m,n,j} ({\mathbf r} )  \left\{ \begin{array}{l}
 \le mn\text{ for $j\le k$} \\
 >mn \text{ for $j>k$} \\
 \end{array} \right. .\nonumber
\end{eqnarray}

\begin{lemma} \label{cut}Let $y=mn$. Further assume that $k\ge 1$ be the maximum $j$ such that ${C}_{m,n,j} ({\mathbf r} ) \le mn$. Then, the achievable diversity-multiplexing tradeoff in
Theorem \ref{thrm:macl1kg1} reduces to:
\begin{eqnarray}
d_{opt}^j = \left\{ \begin{array}{l}
 C_{m,n,j}({\mathbf r})\text{ for $j\le k$} \\
\min(C_{m,n,j}({\mathbf r}),mn+C_{m,n,1}({\mathbf r})) \text{ for
$j=k+1$}\\
 mn+C_{m,n,1}({\mathbf r}) \text{ for $j>k+1$} \\
 \end{array} \right. .\nonumber
\end{eqnarray}
\end{lemma}
\noindent {\bf Proof}: This follows directly from Theorem \ref{thrm:macl1kg1} $\&$ Lemma
\ref{hel}. $\blacksquare$

\begin{figure}[htbp]
\begin{minipage}{4.1cm}
\centering \myincludegraphics[trim=3.1cm 7.5cm 3.3cm 8cm, clip,
height=3.5cm]{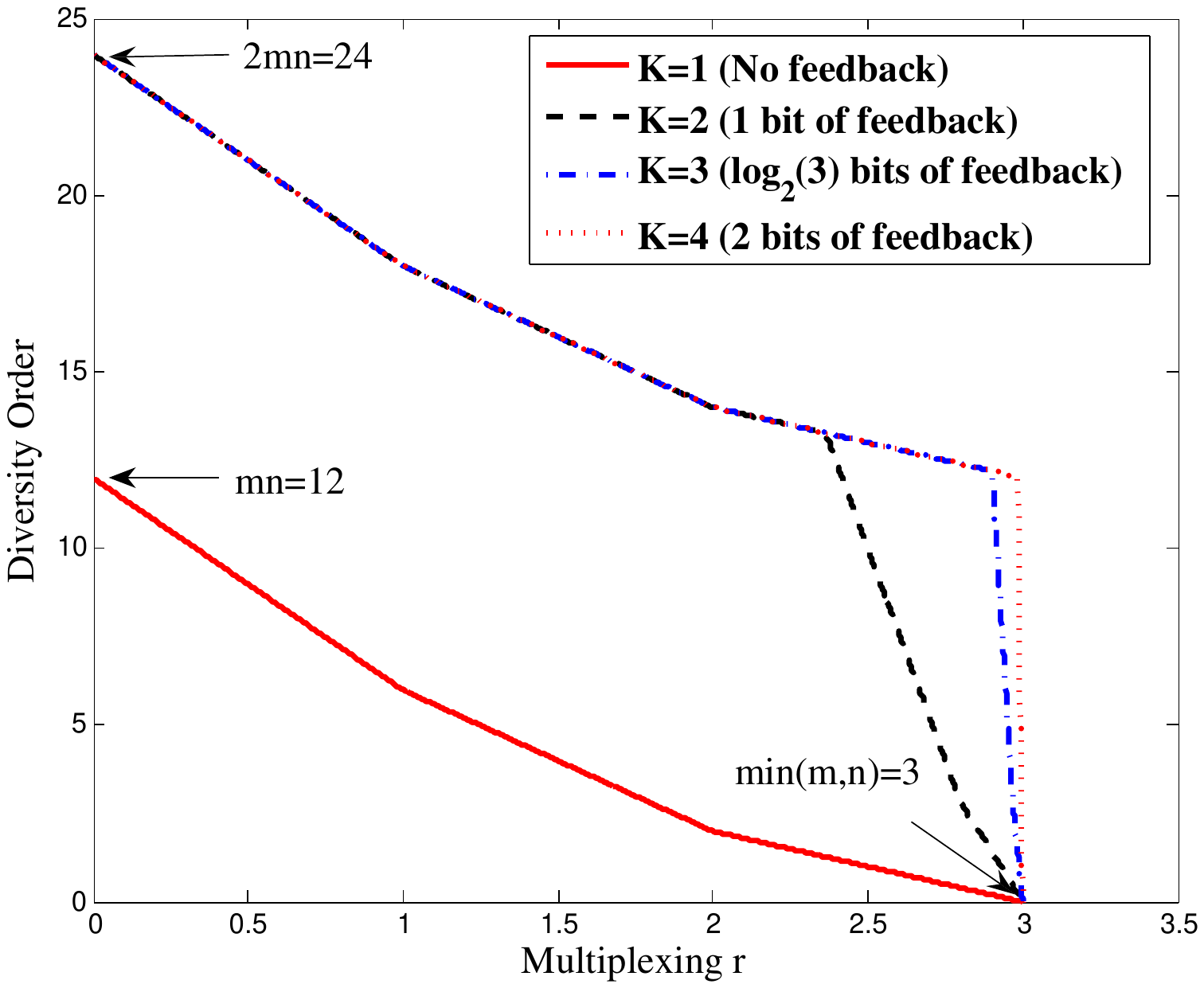} \caption{
Diversity-Multiplexing Tradeoff for MIMO Channel ($m=3, n=4$,
$y=mn$).} \label{figure:mimo1}
\end{minipage}
\begin{minipage}{.4cm}
\end{minipage}
\begin{minipage}{4.1cm}
\centering \myincludegraphics[trim=3.3cm 7.7cm 3.2cm 8.4cm, clip,
height=3.5cm]{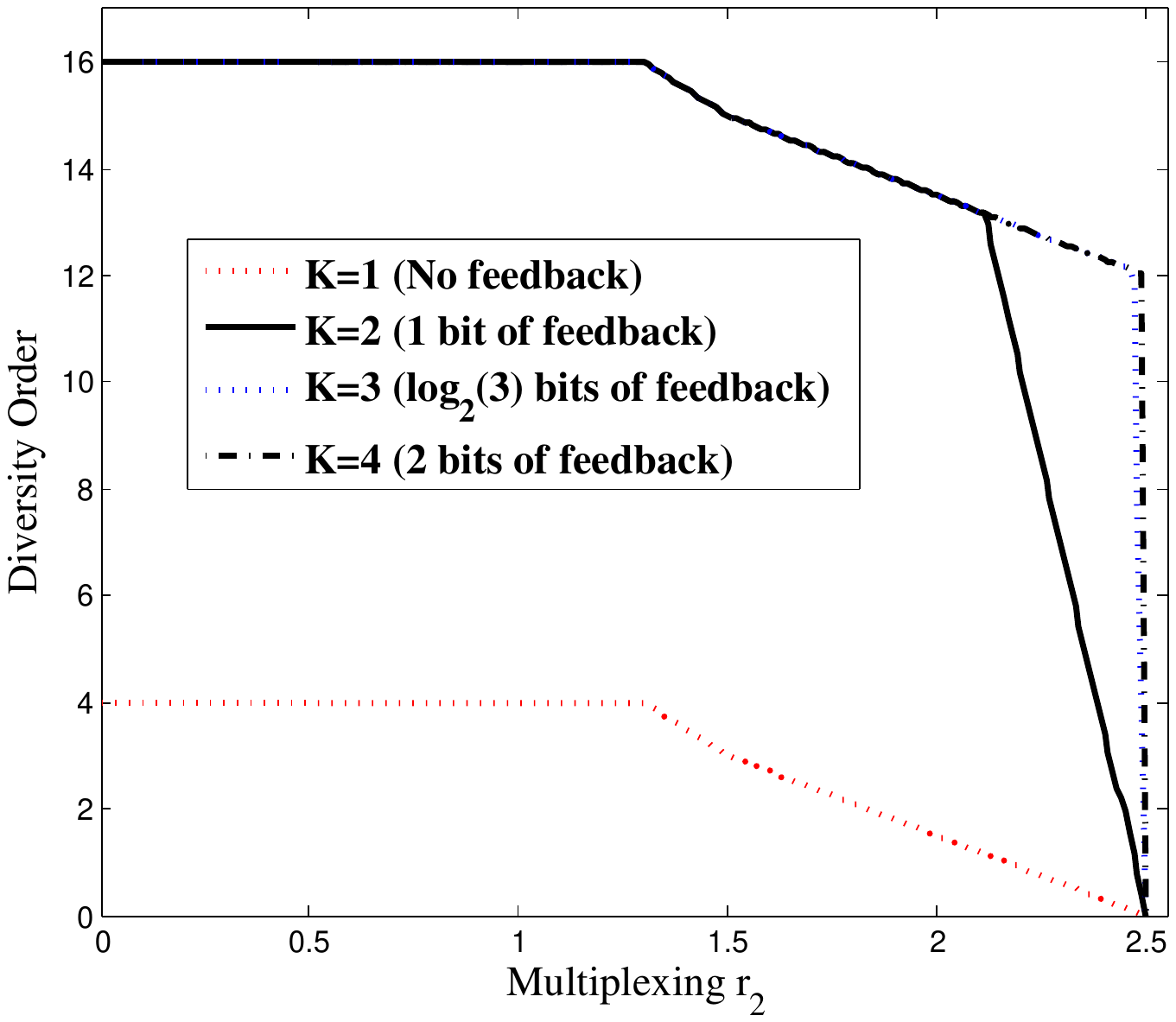} \caption{Diversity-Multiplexing Tradeoff
for MAC ($m=3, n=4$, $y_1=y_2=mn$, $r_1=1.5$).} \label{figure:mac1}
\end{minipage}
\end{figure}

Thus, we find from Lemma \ref{cut} that the diversity increases in
the same fashion with imperfect feedback as it does with perfect
feedback till number of feedback indices $\le k$. For number of
indices $>k+1$, the diversity is limited by $mn + $ the diversity
without feedback. Hence, the gain in diversity order with imperfect feedback over no feedback is limited by $mn$ for any multiplexing gain.
The maximum diversity order that can be achieved with feedback is more than double as compared to the diversity order without feedback for non-zero multiplexing gains
since the diversity order without feedback is less than $mn$. To get this maximum gain in diversity order, $k+2$ feedback indices are sufficient.
Hence, although single bit was enough for multiplexing gains going
to $0$, for general multiplexing gains we need more feedback to
attain maximum diversity. This can also be seen in Figure
\ref{figure:mimo1} for MIMO channel that diversity order of $2mn$
can be achieved with single bit of feedback which is the maximum
possible. Higher amount of feedback indices help at higher
multiplexing to get a gain in diversity order of $mn$ above no-feedback
diversity order. We also see in Figure \ref{figure:mac1} for MAC with two
transmitters ($L=2$) that the diversity order is $mn$ more than the
diversity order without feedback after a certain number of feedback
levels. From these figures, we see that as the multiplexing
increases, more indices are needed to get the maximum gain in diversity order
of $mn$.

\section{Conclusion}
Channel state information at the transmitters is imperfect due to
noise. Inspired by this fact, we constructed a feedback error model
and characterized the diversity multiplexing tradeoff performance
for MAC systems.

\section{Acknowledgement}
This work is supported in part by NSF under
Grants ANI-0338807, CRI-0551692, MRI-0619767 and TF-0635331.


\end{document}